\UseRawInputEncoding
\documentclass[prx,twocolumn,superscriptaddress,floatfix,nopacs]{revtex4}
\usepackage{graphicx,amsfonts,amssymb,amsmath,hyperref,enumerate,footnote}
\usepackage{graphicx}
\usepackage{float}
\usepackage{indentfirst}
\usepackage{verbatim}
\usepackage{tikz}
\usetikzlibrary{quantikz}
\usetikzlibrary{positioning}
\usepackage{array}
\usepackage[utf8]{inputenc}
\usepackage{mathtools}
\usepackage{blindtext}
\usepackage{xcolor}
\usepackage{tikz}
\usetikzlibrary{arrows.meta,
                chains,
                positioning,
                shapes.geometric}
\usepackage{siunitx}
\tikzstyle{startstop} = [rectangle, rounded corners, minimum width=3cm, minimum height=1cm,text centered, draw=black, fill=red!30]
\tikzstyle{io} = [trapezium, trapezium left angle=70, trapezium right angle=110, minimum width=3cm, minimum height=1cm, text centered, draw=black, fill=blue!30]
\tikzstyle{process} = [rectangle, minimum width=3cm, minimum height=1cm, text centered, draw=black, fill=orange!30]
\tikzstyle{decision} = [diamond, minimum width=3cm, minimum height=1cm, text centered, draw=black, fill=green!30]
\tikzstyle{arrow} = [thick,->,>=stealth]

\newif\ifhyper
\hypertrue
\ifhyper
\hypersetup{
   citecolor = {green},
   colorlinks = {true}, 
   urlcolor = {blue} 
}
\fi

\newcommand{\beq}{\begin{equation}}
\newcommand{\eeq}{\end{equation}}
\newcommand{\beqa}{\begin{eqnarray}}
\newcommand{\eeqa}{\end{eqnarray}}

\def\Longarrow{\protect\@lra}
\def\@lra{\relbar\joinrel\relbar\joinrel\relbar\joinrel%
          \relbar\joinrel\rightarrow}

\begin{document}

\title{Improving Gradient Methods via Coordinate Transformations: \\ Applications to Quantum Machine Learning}

\author{Pablo Bermejo}
\affiliation{Multiverse Computing, Paseo de Miram\'on 170, E-20014 San Sebasti\'an, Spain}
\affiliation{Donostia International Physics Center, Paseo Manuel de Lardizabal 4, E-20018 San Sebasti\'an, Spain}

\author{Borja Aizpurua}
\affiliation{Multiverse Computing, Paseo de Miram\'on 170, E-20014 San Sebasti\'an, Spain}

\author{Rom\'an Or\'us}
\affiliation{Multiverse Computing, Paseo de Miram\'on 170, E-20014 San Sebasti\'an, Spain}
\affiliation{Donostia International Physics Center, Paseo Manuel de Lardizabal 4, E-20018 San Sebasti\'an, Spain}
\affiliation{Ikerbasque Foundation for Science, Maria Diaz de Haro 3, E-48013 Bilbao, Spain}

\begin{abstract} 

{In this paper we introduce a generic strategy to accelerate and improve the overall performance of machine learning methods in order to escape from barren plateaus.}
Machine learning algorithms, both in their classical and quantum versions, heavily rely on optimization algorithms based on gradients, such as gradient descent and alike. The overall performance is dependent on the appearance of local minima and barren plateaus, which slow-down calculations and lead to non-optimal solutions. In practice, this results in dramatic computational and energy costs for AI applications.  Our {novel} method is based on coordinate transformations, somehow similar to variational rotations, adding extra directions in parameter space that depend on the cost function itself, and which allow to explore the configuration landscape more efficiently. The validity of our method is benchmarked by boosting a number of quantum machine learning algorithms, getting a very significant improvement in their performance. 

\end{abstract}

\maketitle

\section{Introduction}

Machine learning is revolutionizing society. We have witnessed it recently with the advent of game-changing applications such as ChatGPT, where Large Language Models \cite{LLM} allow for unprecedented human-computer interaction. Such systems have a neural structure at their roots, with weights that must be optimized so as to minimize some error cost function. This optimization is usually done via gradient methods such as gradient descent, stochastic gradient descent, adaptive moment estimation, and alike, which are not free from problems such as barren plateaus and local minima. One important consequence is that current artificial intelligence (AI)  systems suffer from long and complex training procedures, amounting to dramatic and unsustainable computational and energy costs \cite{refEconomist}. And given the increasingly-huge demand of AI systems, the situation is even worse: we are in big need of more efficient machine learning.

Mathematically speaking, numerical algorithms used for optimization problems are mostly based on techniques that sweep the whole hyperspace of solutions. This landscape might present a tractable shape where the optimal solution can be easily found, as in convex problems. However, most interesting problems have an ill-defined landscape of solutions, with non-predictable shapes plenty of complexities impeding analytical and numerical methods to properly act on them. Gradient methods explore this landscape by computing local gradients of the cost function and updating the parameters according to the computed local slopes. More complex optimization methods, such as the widely-used stochastic gradient descent and adam optimizer, are based on the same idea. 

The work that we present here addresses the two main caveats of employing gradient methods for optimization purposes: local minima and barren plateaus. These features emerge inherently due to the shape of the cost function. In fact, both limitations can be understood as coming from moving along certain directions in the landscape of solutions. Changes in the optimization variables are the result of changes in the cost function, as the variables are modified. This is evidently an obstacle when we find a flat region in the landscape of the cost function. 

In this paper we propose an alternative way to navigate the landscape of solutions by introducing extra freedom in the directions of the parameters' update. This is done by using (i) changes of coordinates, and (ii) adding extra dimensions related to the cost itself. In our specific implementation, we consider changes to hyperspherical coordinates, as well as frame rotations. In order to do that, we treat the cost as an extra variable to be optimized, and implement a self-consistent variational method in which the cost axis not only represents the value to be optimized, but also serves as an extra dimension to escape from local minima and barren plateaus. {Importantly, we stress that our approach aims not to lower the cost function values, but rather and specifically to move out from barren plateaus.}

The structure of this paper is as follows. In Sec.\ref{sec2} we present our boosting methodology via coordinate transformations. In Sec.\ref{sec3} we show a set of benchmarks to validate our idea, where we systematically improve the performance of a number of well-known quantum machine learning algorithms. Finally, in Sec.\ref{sec4} we present our conclusions and discuss future directions. 

\begin{figure}
\centering
  \includegraphics[width=0.69\columnwidth]{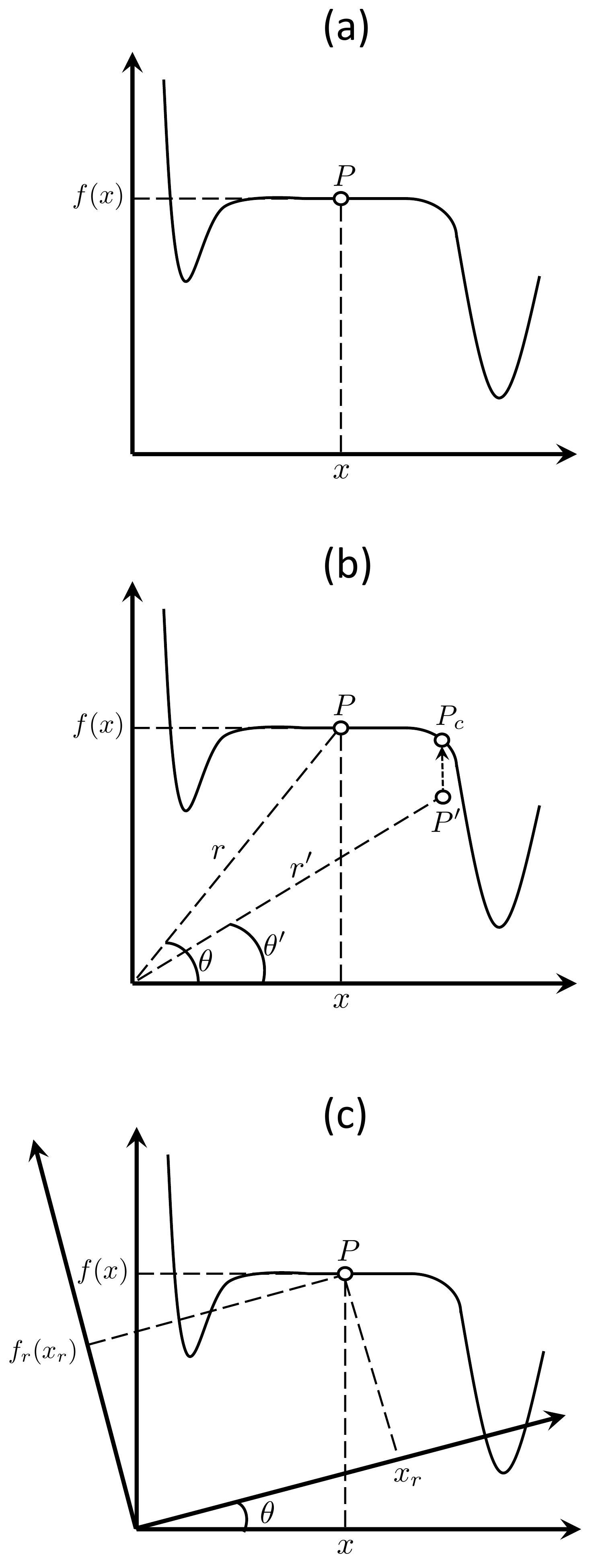}
  \caption {(a) Cost function $f(x)$ presents a plateau as a function of optimization variable $x$, so that gradient methods get stalled at point $P$ due to a null gradient. (b) A change in the polar coordinates of point $P$ leads to a point $P'$, which can then be ``collapsed" back to the landscape of the cost function, leading to point $P_c$. Optimization from point $P_c$ is no longer stalled, since the gradient is non-zero. (c) A frame rotation leads to a description of point $P$ with different cartesian coordinates $f_r(x_r)$ and $x_r$. In the new rotated frame, the gradient at $P$ is non-zero, so that gradient methods are no longer stalled.}
  \label{fig1}
\end{figure}

\section{Methodology}
\label{sec2}

An optimization problem can be stated as the minimization of a scalar real cost function $f(\vec{x})$ of $n$ variables $\vec{x} \equiv (x_1, x_2, ... , x_n)$. Methods based on gradient descent apply an update of parameters mainly based on the calculation of changes in the function when there is a change in the coordinate values, i.e., 
\beq
\Delta f(\vec{x}) = \left(\frac{\partial f}{x_1}, \frac{\partial f}{x_2}, ... , \frac{\partial f}{x_n} \right)\Bigg|_{\vec{x}_0}
\cdot \Delta \vec{x} = \vec{\nabla} f(\vec{x})|_{\vec{x}_0}
 \cdot \Delta \vec{x}, 
\eeq
where $\Delta \vec{x} = \vec{x}_1 - \vec{x}_0$ for some arbitrary $\vec{x}_1$, and where $\vec{x}_0$ is the point in parameter space where the gradient $\vec{\nabla} f(\vec{x})$ of the function is computed. As is clear from this equation, a gradient close to zero represents flat regions in the landscape of solutions, implying no updates in the parameters, and therefore disabling further optimization steps, so that the optimization gets stuck.

To avoid that, we propose a new method based on a change of coordinates. Our idea can be well understood by considering the case of optimization of a one-dimensional cost function, represented by the landscape in Fig. \ref{fig1}. In Fig. \ref{fig1}a, the cost function has a plateau and therefore the gradient is zero, so that optimization is stalled. However, in Fig. \ref{fig1}b we see that we can avoid this by \emph{changing the polar coordinates of point $P$ in the landscape}, i.e., in the two-dimensional plane formed by the axis corresponding to the coordinate $x$ to be optimized \emph{and} the cost function $f(x)$. The new point $P'$ can then be ``collapsed" back to the landscape of the cost function, so that we can escape from the null-gradient region. Alternatively, we can also rotate the frame, as in Fig. \ref{fig1}c. In the new rotated frame, the rotated cost function $f_r(x_r)$ does not present a null gradient with respect to the rotated parameter $x_r$, so that moving by gradient methods is again possible in the new frame.

While the above is a simple picture explaining the basic idea of our algorithm, it captures many of its basic ingredients. Long story short: if the problem coordinates do not work, then we change them by including the cost as an extra coordinate. This change of coordinates can be a rotation, or a more generic one. 

Let us now explain how to implement our method in full generality. As sketched above, there are two versions of it: changing to hyperspherical coordinates, or rotating the frame. As we shall see, both approaches are similar, though not equivalent. Let us start with the change to hyperspherical coordinates.

\subsection{Option 1: hyperspherical coordinates}
\label{active point}
Consider the cost function $f(\vec{x})$ and the original $n$ cartesian coordinates $\vec{x}$. In the landscape space, this defines a point $P$ in the $(n+1)$-dimensional space described by the coordinates \emph{and} the cost function: 
\begin{equation}
P = \left[ x_1, x_2, \cdots, x_n, f(\vec{x}) \right]. 
\label{point}
\end{equation}
The procedure to follow in one iteration step is as follows: 
\begin{enumerate}
    \item Make a change of coordinates. Without loss of generality, we will use ($n+1$)-dimensional hyperspherical coordinates. Therefore, make the change 
    \begin{equation} 
    P = \left[ x_1, x_2, \cdots, x_n, f(\vec{x}) \right] \rightarrow P = \left[ \theta_1, \theta_2, \cdots, \theta_n, r \right], 
    \end{equation}
    with $\{ \vec{\theta}, r \}$ being the $n+1$ hyperespherical coordinates for point $P$. 

    \item\label{grad_descent_active} Perform gradient descent on the new set $\{\vec{\theta},r\}$ with reference cost value $f(\vec{x})$. In practice, we will simply update the variables in the hyperspherical coordinates using changes of the cost function in the original cartesian description (the one we want to minimize) each time we update these parameters. We can do so by projecting each point in the hyperspherical description into the cartesian cost function to obtain that reference value.
    
    This translates into an iterative switiching process from one description to the other in orther to be able to move in the hyperspherical coordinates while the changes in the original cost function are the ones driving the parameter updates.

    This procedure will end up giving us a new point in the ($n+1$) dimensional plane, defined by the change $\{\vec{\theta},r\} \rightarrow \{\vec{\theta}',r'\}$ after all the parameters have been updated:  
    \begin{equation}
    P' = \left[ \theta'_1, \theta'_2, \cdots, \theta'_n, r' \right]. 
    \end{equation}
 
     \item Once the gradient descent is completed, we can  make a final transformation $\{\vec{\theta}',r'\} \rightarrow \{\vec{x_{\rm c}},f_c(\vec{x_{\rm c}})\}$, retrieving the point P$_{\rm c}$ which is just defined by the projection of point P´ into the original cost function:
    \begin{equation}
     P_{{\rm c}} \equiv \left[ x'_1, x'_2, \cdots, x'_n, f'(\vec{x}') \right], 
     \end{equation}
     where $c$ stands for "collapsed".

\end{enumerate}
These steps are to be repeated until convergence, as in usual methods based on gradient descent, with the difference that the gradient is computed via the change of coordinates. Additionally, this change of coordinates can be implemented only when we fall in local minima or barren plateaus, in order to move out from the region, or alternate between the two during the optimization.

\begin{table*}
\hspace{-0.8cm}
\setlength{\extrarowheight}{4pt}
\centering
\begin{tabular}{|c|c|c|}
\hline
\textbf{Paper}  & ~~\textbf{Original} ~~& ~~\textbf{Hyperspherical}~~ \\  \hline   \hline
\begin{tabular}[c]{@{}c@{}}Alleviating Barren Plateaus \cite{Cost_function_dependent}\end{tabular} &  29.14          & 4.7                \\ \hline
\begin{tabular}[c]{@{}c@{}}Accelerating VQEs with Quantum Natural \\ ~~Gradient: Single-qubit VQE / Hydrogen VQE (Adam) \cite{variational_eigenvalue_solver}~~\end{tabular} & 182 / 103          & 160 / 62                 \\ \hline
\begin{tabular}[c]{@{}c@{}}Function Fitting Using Quantum Signal \\Processing - Polynomial \cite{function_fitting_QSP}\end{tabular}   & 90            &   54                  \\ \hline
\begin{tabular}[c]{@{}c@{}}Variational Classifier Iris Classification \cite{classification_qnn} \end{tabular}                                              &      100            &    30                \\ \hline
\begin{tabular}[c]{@{}c@{}}Variational Quantum Thermalizer \cite{variational_quantum_thermalizer}  \end{tabular}                                                         &       103           &               51         \\ \hline
\end{tabular}
\caption{Advantage in number of iterations for a number of algorithms, with respect to the original implementation using Xanadu's PennyLane, and where the new implementation uses the change to hyperspherical coordinates. Numerical data is the average number of iterations.}

\label{table:results_hyperspherical}
\end{table*}

\subsection{Option 2: plane rotations}

An alternative to the change to hyperspherical coordinates is to rotate the frame. Consider again the cost function $f(\vec{x})$ and the original $n$ cartesian coordinates $\vec{x}$. Following Eq.(\ref{point}), this generates a point $P$ in the $(n+1)$-dimensional space defined by the coordinates and the cost function. This point can also be described as a vector $\vec{P}$ in the considered frame. The procedure to follow this time in one iteration step is then as follows: 
\begin{enumerate}
    \item Make a change of coordinates in the frame. Without loss of generality, we will use a $(n+1)$-dimensional rotation of the axis.  
 \begin{equation} 
    \vec{P} = \left( x_1, \cdots, x_n, f(\vec{x}) \right) \rightarrow  \vec{P}_{r} = R \cdot \vec{P}
\end{equation}
 with $R$ the $(n+1)$-dimensional axis rotation and subscript $r$ standing for rotated. A convenient simplification is to implement a $2$-dimensional rotation of the plane formed by some individual coordinate $x_i$ and the function $f(\vec{x})$, i.e., 
\begin{equation}
\begin{aligned}
\vec{P} = \left( x_1, \cdots, x_i, \cdots, x_n, f(\vec{x}) \right) \rightarrow \\ 
\vec{P}_{r} = R_2 \cdot \vec{P} = \left( x_1, \cdots, x_{ir}, \cdots, x_n, f_r(\vec{x_r}) \right)
\end{aligned}
\end{equation} 
where subindex $r$ indicates the rotated variables, and with $R_2 \in SO(2)$ the $2 \times 2$ matrix corresponding to a rotation of the $(x_i, f(\vec{x}))$ plane. This is a good choice if we think that there may be a barren plateau or a local minima in the direction of coordinate $x_i$. Additionally, the rotation angle may not be necessarily fixed, so that it can work as an extra hyperparameter in the optimization, which has been the case for most of our results. For simplicity, and without loss of generality, from now on we assume such a 2d rotation. 

\item In the rotated frame, perform a similar procedure to the one explained in point \ref{grad_descent_active} from subsection \ref{active point} to compute the gradient descent. Now, instead of switching from one description to the other by means of projections from the updated parameters to the original cost function, we will simply move from one to the other by means of relative rotation matrices.

This will lead us to a final point, after all the updates have been performed, given by the inverse rotation R$_2^{-1}$,
recovering a new updated cost value $f_u(\vec{x}_u)$:

         \begin{equation}
   \vec{P_{u}} =  R_2^{-1} \cdot \vec{P}_{{r}}'  =  \left( x_{1u}, x_{2u}, \cdots, x_{nu}, f_{u}(\vec{x}_{u}) \right) 
    \end{equation}
    \end{enumerate}

Notice, that we use subindex u to make explicit the fact that now the change of coordinates does not come from a projection, so there is no collapsed cost function, but  is instead directly obtained from relative matrix rotations.

Again, these steps are to be repeated until convergence, as in usual methods based on gradient descent. Furthermore, one could implement the rotations just when falling in local minima or barren plateaus, in order to move out from the region, or alternate between the two frames during the optimization.

{It is also worth noticing that placing the rotating point elsewhere does not affect the performance of the method. All in all, computing the gradients with respect to any other point in the parameter space while retaining description of parameters with respect to the original should not give either any improvement.}

\bigskip 

The two methods presented here make use of a higher-dimensional space, where the cost function is included as an extra dimension, and is optimized self-consistently. One may be tempted to say that all the advantage of our method comes from extending the variables to a higher-dimensional space. While this is tempting, it is actually not true. We have also tested implementations where we just extended the coordinate space, but without adding the cost function as an extra coordinate, in a way similar to kernel methods. In such implementations, we have seen no significant advantage in convergence, in stark contrast with the two methods presented above. We therefore conclude that including the cost function as an extra coordinate to be optimized self-consistently is in fact fundamental. {We want to remark that, physically, in the parameter space, we are not introducing any new dimension. The parameter landscape is built upon a number of variables which are combined in a certain manner to define a cost function. So evaluating these parameters at a given point gives some cost value. We are proposing to use that same cost function, which has support on a given parameter space, to add one more dimension through which we can navigate to optimize the parameter search. 

Concerning the learning rate, upon applying coordinate transformations to escape these plateaus, the optimization landscape undergoes a significant change. The gradients' scale can be substantially different in the new coordinates, which implies that the previously optimal learning rate may no longer be effective. A higher learning rate, in theory, can indeed introduce a greater degree of perturbation in the parameters, facilitating the escape from a plateau. However, this must be balanced carefully; too high a learning rate can lead to overshooting minima or increased oscillations in the training process, potentially destabilizing convergence. This is also aligned with the fact that learning rate is a problem-dependent parameter. It usually requires a preliminary grid search to find the optimal learning rate for each use case. Here, we are altering the optimization process itself so, in the same way learning rate is problem-dependent, one can think that any modification in the optimization process will require further tuning of important hyperparameters as the learning rate.}

In what follows we show the benefits of our approach by implementing a number of benchmarks.

\section{Results}
\label{sec3}

For the purpose of implementation and testing, a set of quantum machine learning algorithms has been used for the sake of comparison, with the original implementation being done via PennyLane \cite{pennylane}. Some of these algorithms introduce new methods for solving optimization processes, as it is the case of quantum natural gradient descent \cite{quantum_natural_gradient}, or the construction of local cost functions in \cite{Cost_function_dependent}. Some other methods propose optimization algorithms for solving certain tasks. 

Here we show how our algorithm actually boosts the performance of all the tested methods. For the purpose of the analysis, we have revisited 18 different algorithms, claiming significant advantages (i.e., faster convergence or lower cost function values) for 15 of them. Let us also stress that this improvement is obtained against optimal and fine-tuned implementations of the original algorithms using PennyLane, and not merely naive ones. This also accounts for using the most efficient optimizer for each case, which happens to be mostly Adam optimizer.

Let us start by showing in Table \ref{table:results_hyperspherical} a summary of our results for the method using the change to hyperspherical coordinates. As we see in the table, the change to hyperspherical allows for substantially-faster convergence in five quantum machine learning algorithms, with respect to their original optimal implementation. Let us now describe four of these examples.   

\begin{figure}
  \centering
  \includegraphics[width=\columnwidth]{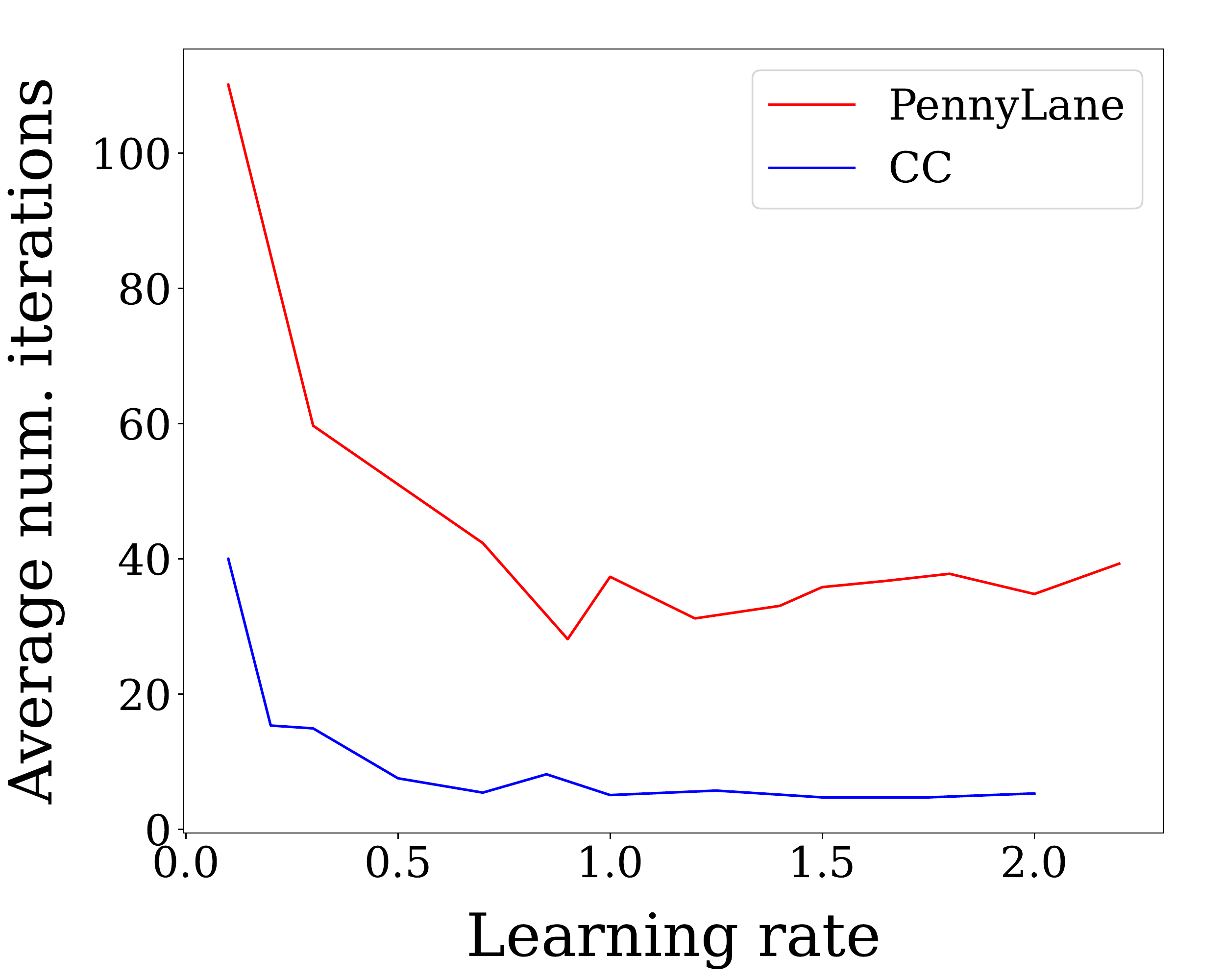}
  \caption{[Color online] Improvement in the average number of iterations as a function of the learning rate, for the algorithm in Ref. \cite{Cost_function_dependent} to alleviate barren plateaus with local cost functions. Comparison between PennyLane and the Change of Coordinates (CC) implementations.}
  \label{Alleviting}
\end{figure}
\begin{figure}
  \centering
  \includegraphics[width=\columnwidth]{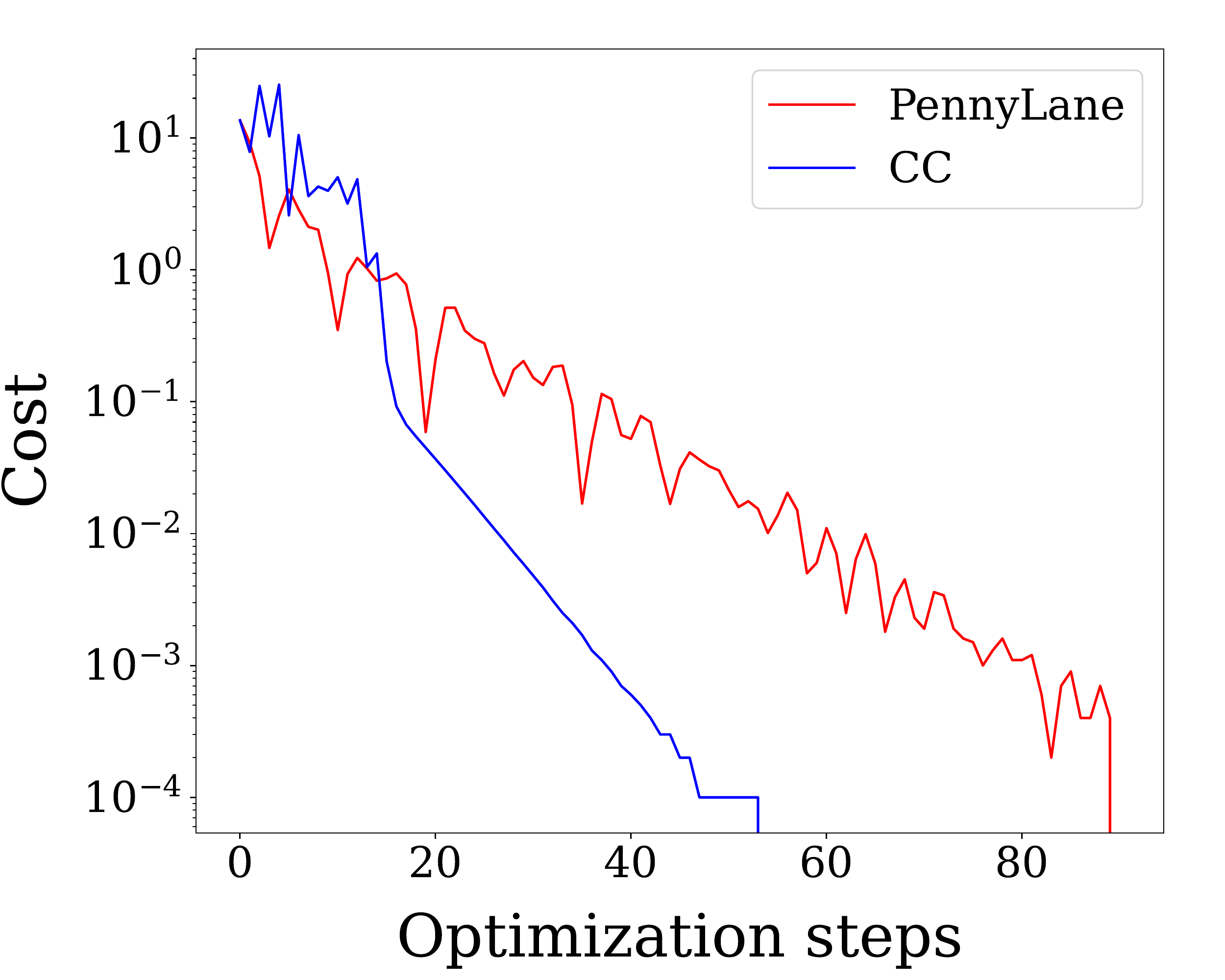}
  \caption {[Color online] Convergence of cost function versus number of optimization steps, in function fitting using a quantum signal processing polynomial, for the algorithm in Ref. \cite{function_fitting_QSP}. Comparison between PennyLane and the Change of Coordinates (CC) implementations.}
  \label{Function fitting}
\end{figure}
\begin{figure}
\centering
  \includegraphics[width=\columnwidth]{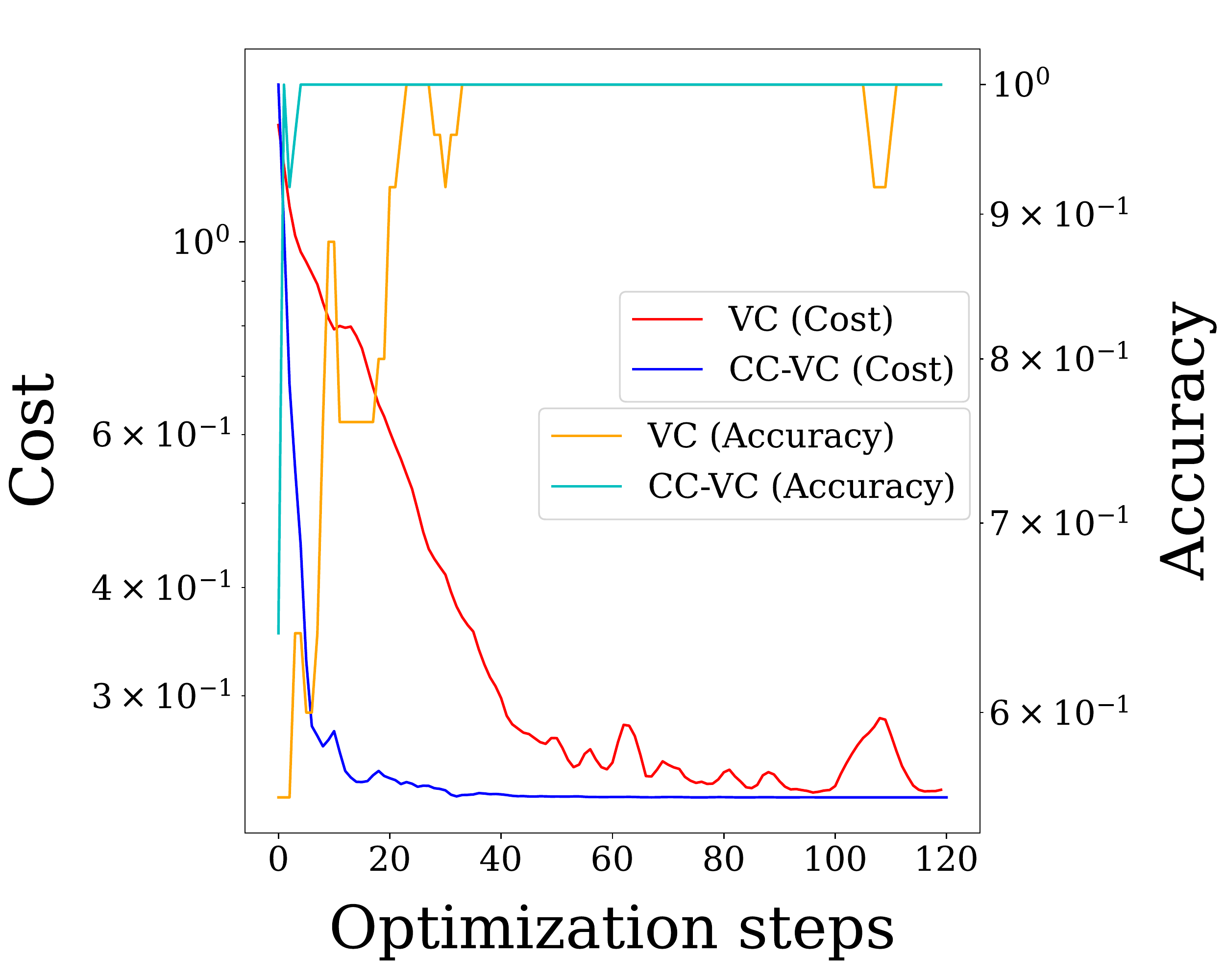}
  \caption {[Color online] Convergence of cost function and accuracy versus number of optimization steps, in the variational quantum classifier from Ref. \cite{classification_qnn} for the Iris dataset. Comparison between PennyLane and the Change of Coordinates (CC) implementations.}
  \label{VQC}
\end{figure}
\begin{figure}
    \centering
    \includegraphics[width=\columnwidth]{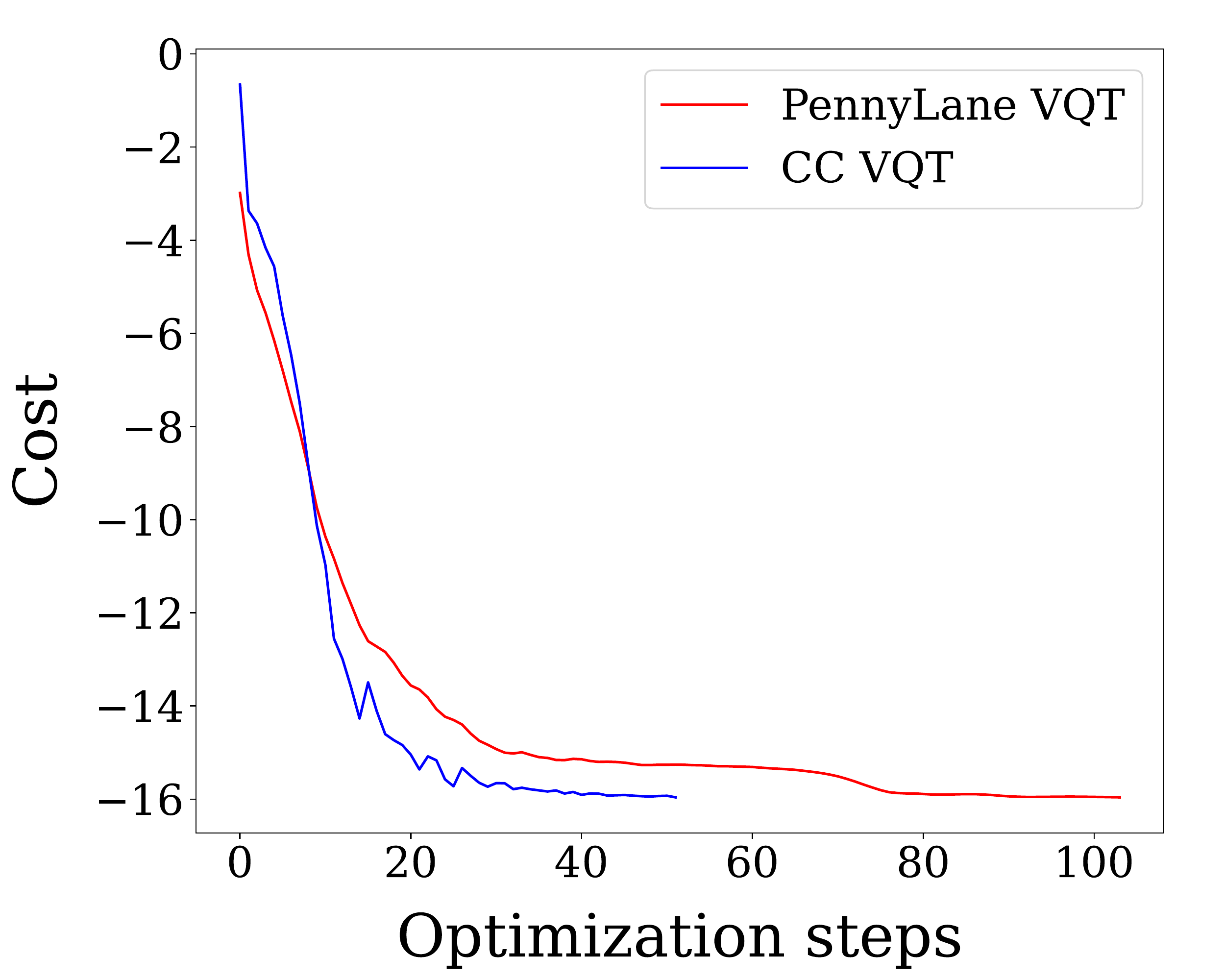}
    \caption {[Color online] Convergence of cost function versus number of optimization steps, in the variational quantum thermalizer from Ref. \cite{variational_quantum_thermalizer}. Comparison between PennyLane and the Change of Coordinates (CC) implementations.}
    \label{fig:vqt}
\end{figure}

In Ref. \cite{Cost_function_dependent}, it is shown that local cost functions with a shallow quantum circuit improve the trainability of variational quantum algorithms. Global cost functions show barren plateaus that cannot be overcome by usual means, whereas local cost functions improve the convergence. In this case, we implemented simulations for a set of 50 different initial configurations to solve a certain optimization problem.  The original global approach showed 30$\%$ success in overcoming barren plateaus with more than 100 iterations in average for convergence. Local cost functions showed a ratio of success of 98$\%$ with an average number of iterations equal to 29.14. And, by using our approach with hyperspherical coordinates, we reach a success of 100$\%$ with 4.7 iterations in average, implying a very significant improvement, see Fig. \ref{Alleviting}.

Next, in Ref. \cite{function_fitting_QSP} it was considered the idea of quantum signal processing in order to perform an optimization process for function fitting. Again, we have observed that our implementation using hyperspherical coordinates provides a speed up in optimization steps for a faster convergence to lower values of the cost. In particular, to reach a limiting cost value of $10^{-5}$, we were able to reduce the number of iterations for convergence from 90 in the original implementation to 54 in ours, this is, a reduction of 40\% of iterations, see Fig. \ref{Function fitting}.

\begin{table*}
\hspace{-0.8cm}
\setlength{\extrarowheight}{4pt}
\begin{tabular}{|c|c|c|}
\hline
\textbf{Paper}                                                                                                              & ~~\textbf{PennyLane}~~ & ~~\textbf{Rotational Axis}~~ \\  \hline \hline
\begin{tabular}[c]{@{}c@{}}Alleviating Barren Plateaus  \cite{Cost_function_dependent}\end{tabular} &  49-1/29.14          & 50-0/24.92                 \\ \hline
\begin{tabular}[c]{@{}c@{}} Quantum natural gradient  \\
(QNG / Adam) and (GD /Adam) \cite{quantum_natural_gradient}\end{tabular}                                                                                               & 26 / 88                 & 19 /75                       \\ \hline
\begin{tabular}[c]{@{}c@{}}Accelerating VQEs with Quantum Natural \\ Gradient: Hydrogen VQE (QNG / GD) and (GD) \cite{variational_eigenvalue_solver} \end{tabular} & 20 / 66          & 16                 \\ \hline
\begin{tabular}[c]{@{}c@{}}Accelerating VQEs with Quantum Natural \\ ~~Gradient: Single-qubit VQE / Hydrogen VQE (Adam)~~ \cite{variational_eigenvalue_solver} \end{tabular} & 182 / 103          & 158 / 98                 \\ \hline

\begin{tabular}[c]{@{}c@{}}Function Fitting using Quantum Signal \\Processing - Polynomial / Non-Polynomial \cite{function_fitting_QSP}\end{tabular}   & 90 / 112           & 74 / 94                  \\ \hline
Variational Quantum Thermalizer \cite{variational_quantum_thermalizer}                                                                                             & 103                & 51                      \\ \hline
\begin{tabular}[c]{@{}c@{}}Function Fitting with Photonic \\ Quantum Neural Network \cite{cv_qnn}\end{tabular}                            & 70                 & 51                       \\ \hline
\begin{tabular}[c]{@{}c@{}}Perturbative Gadgets for Variational \\ Quantum Algorithms \cite{perturbative_gadget_delay_onset}\end{tabular}                          & 140                & 110                      \\ \hline
Variational Classifier Iris classification \cite{classification_qnn}                                                                                 & 96                 & 80                       \\ \hline

VQE in Different Spin Sectors S=0 / S=1 \cite{variational_eigenvalue_solver}                                                                                    & 166 / 54           & 142 / 34                 \\ \hline

\end{tabular}
\caption{Advantage in number of iterations for a number of algorithms, with respect to the original implementation using Xanadu's PennyLane, and where the new implementation uses frame rotations. Numerical data is the average number of iterations.}

\label{table:results2}
\end{table*}

\begin{table*}
\hspace{-0.8cm}
\setlength{\extrarowheight}{4pt}
\begin{tabular}{|c|c|c|}
\hline
\textbf{Paper}                                                                                                              & ~~\textbf{PennyLane}~~ & ~~\textbf{Rotational Axis} ~~\\  \hline \hline 
VQCO: 8 qubits, 80 layers    \cite{cv_optimizer} & $1.384\times10^{-4}$ & $6.942\times 10^{-5}$     \\ \hline
Quantum models as Fourier series \cite{data_encoding}         & $9.67\times10^{-13}$            & $1.55\times10^{-13}$                \\ \hline
~~Variational Classifier Iris classification \cite{classification_qnn}  ~~   & 0.2323              & 0.2303               \\ \hline
Variational Quantum Linear Solver   \cite{variational_quantum_linear_solver}       & $5.83\times10^{-10}$ & $1.16\times10^{-10}$     \\ \hline

\end{tabular}
\caption{Advantage in cost function value for a number of algorithms, with respect to the original implementation using Xanadu's PennyLane, and where the new implementation uses frame rotations. Numerical data is the average number of iterations.}
\label{table:results3}
\end{table*}

Additionally, in Ref. \cite{classification_qnn} the authors proposed a quantum neural network based on a variational quantum circuit to perform classification of a data set. Here we plot in Fig. \ref{VQC} the convergence of the classification procedure for a specific quantum circuit configuration (namely, a fixed ansatz with the same number of qubits and layers) to classify the Iris dataset. The plot shows that our approach with hyperspherical coordinates improves significantly the convergence stability of the algorithm while reducing the number of steps to achieve convergence, both in the accuracy of the classification as well as in the cost function to be optimized. For instance, to achieve a similar value of the cost of $\approx 0.23$, we reduce the number of steps from roughly $100$ to roughly $30$, implying a speedup of 70\% with the new implementation. From this example we also see that our new implementation provides a smoother convergence. Avoiding randomness and drastic variations in the optimizing curve can also become a concern for certain applications, and we understand this as a sign that our implementation fits better the problem landscape.

As a last example for the hyperspherical approach, we consider the algorithm in Ref.  \cite{variational_quantum_thermalizer}. Here the authors proposed a generalization of the variational quantum eigensolver to compute thermal states via a hybrid optimization procedure. Again, as shown in Fig. \ref{fig:vqt}, we can reduce the number of optimization steps to find the thermal states. In particular, we reach convergence in almost half the number of optimization steps, reducing from 103 in the original implementation to just 51 in ours.

\begin{figure}
\centering
    \includegraphics[width=\columnwidth]{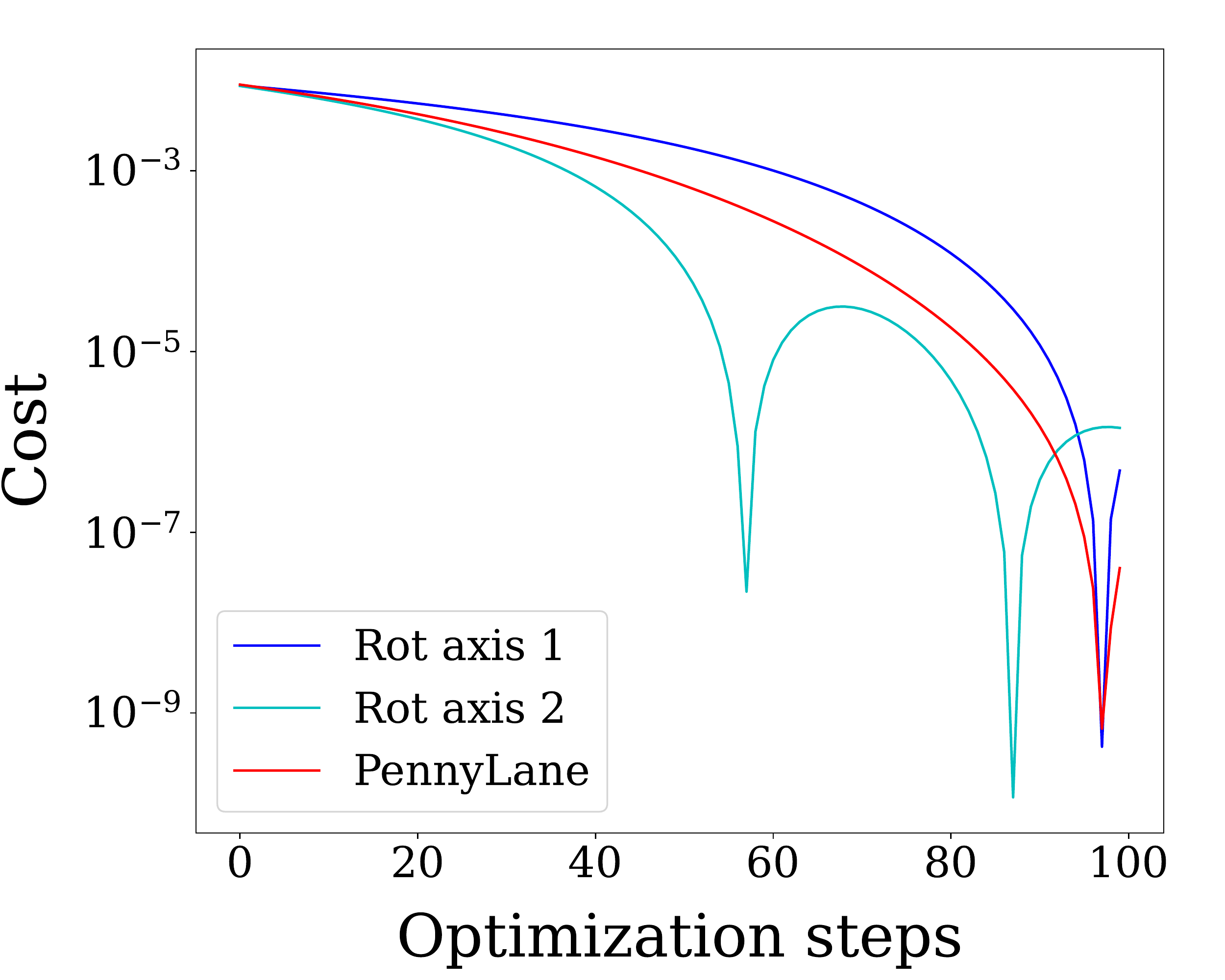}
        \caption {[Color online] Convergence of cost function versus number of optimization steps, in the variational quantum linear solver from Ref. \cite{variational_quantum_linear_solver}. Comparison between PennyLane and the rotation implementations, where we implemented two rotation approaches.}
    \label{fig:linear}
\end{figure}
\begin{figure}
    \centering
    \includegraphics[width=\columnwidth]{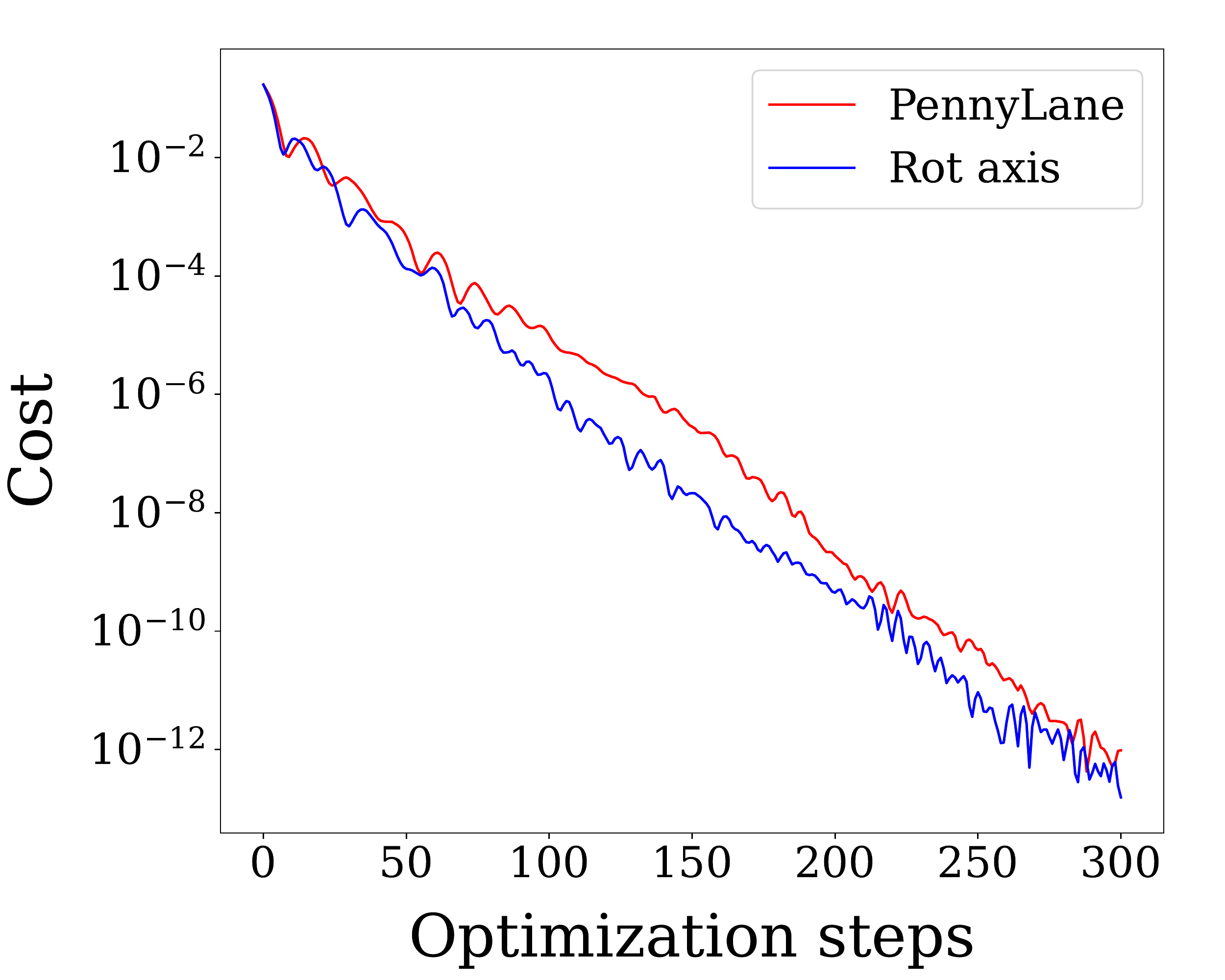}
 \caption {[Color online] Convergence of cost function versus number of optimization steps, in the quantum models for Fourier series from Ref. \cite{data_encoding}. Comparison between PennyLane and the rotation implementations.}    \label{fig:fourier}
\end{figure}

Next, let us consider our second approach, namely the frame rotations. Our results are summarized in Tables \ref{table:results2} and \ref{table:results3}, showing examples of improvement in the number of iterations and in the cost function value, respectively. The method of frame rotations allows for substantially-faster convergence in ten quantum machine learning algorithms and lower cost functions in three, with respect to their original optimal implementation. Let us briefly sketch three of these examples.  

In Ref. \cite{variational_quantum_linear_solver}, the authors proposed a novel algorithm  to solve linear systems of equations by using a hybrid quantum-classical approach suitable for near-term quantum devices. We have compared this approach with two different configurations of our rotation method, where rotations were implemented for different planes. As shown in Fig. \ref{fig:linear}, depending on the rotated plane we can achieve faster convergence as well as lower values of the cost function, as compared to the original approach.  

Next, the work in Ref. \cite{data_encoding} describes an optimization process built in order to test the expressive power of specific quantum circuits and their ability to fit certain functions by learning features defining their Fourier series. Our comparison is shown in Fig. \ref{fig:fourier}. Even if the final cost is tiny for both cases, we can observe in the plot that the computed cost function in our approach is almost always lower than the one computed with the original method. 

\begin{figure}
    \centering
    \includegraphics[width=\columnwidth]{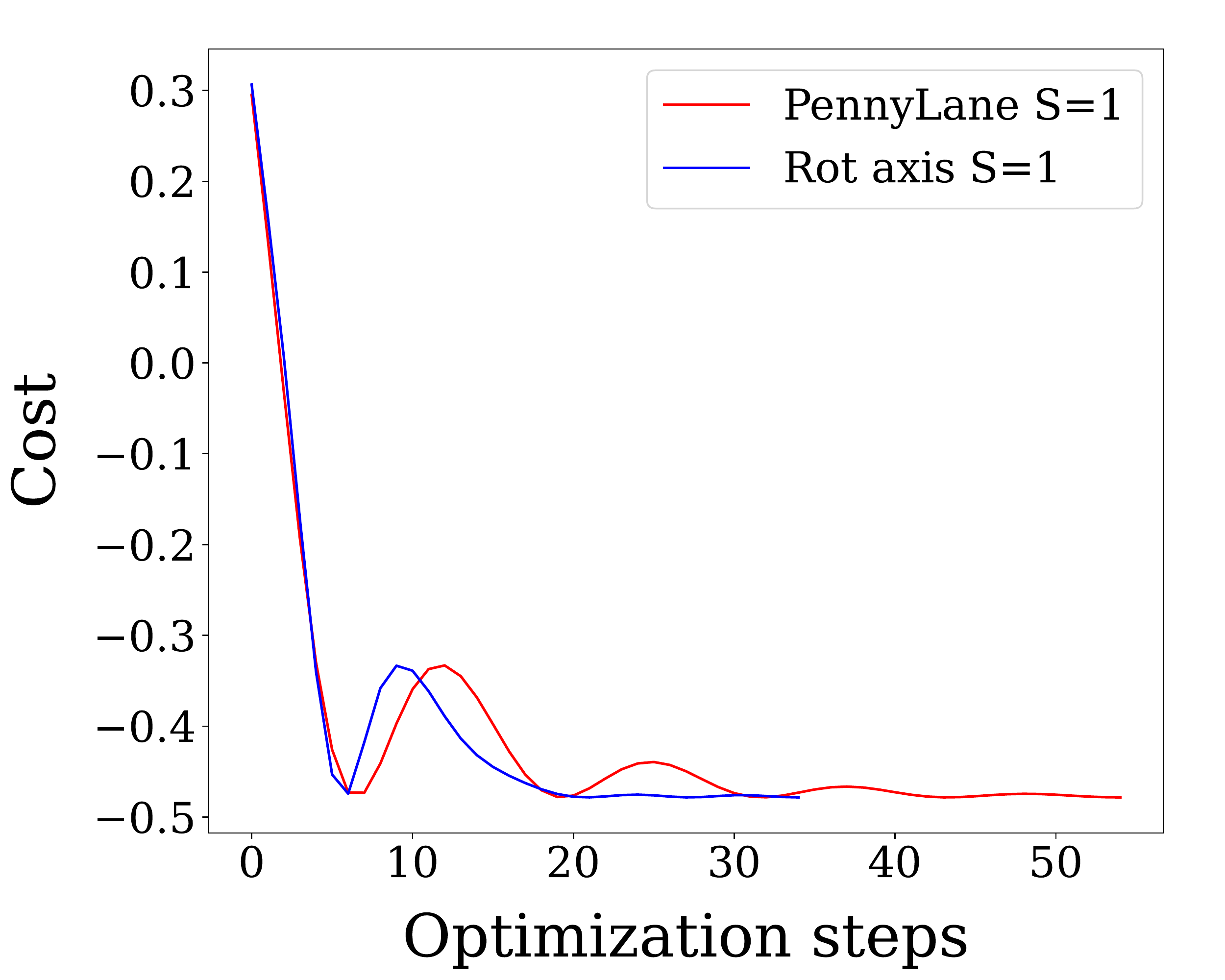}
     \caption {[Color online] Convergence of cost function versus number of optimization steps, in the variational quantum eigensolver for different spin sectors from Ref. \cite{variational_eigenvalue_solver}. Comparison between PennyLane and the rotation implementations.}   
    \label{fig:Spins}
\end{figure}
\begin{figure}
    \centering
    \includegraphics[width=\columnwidth]{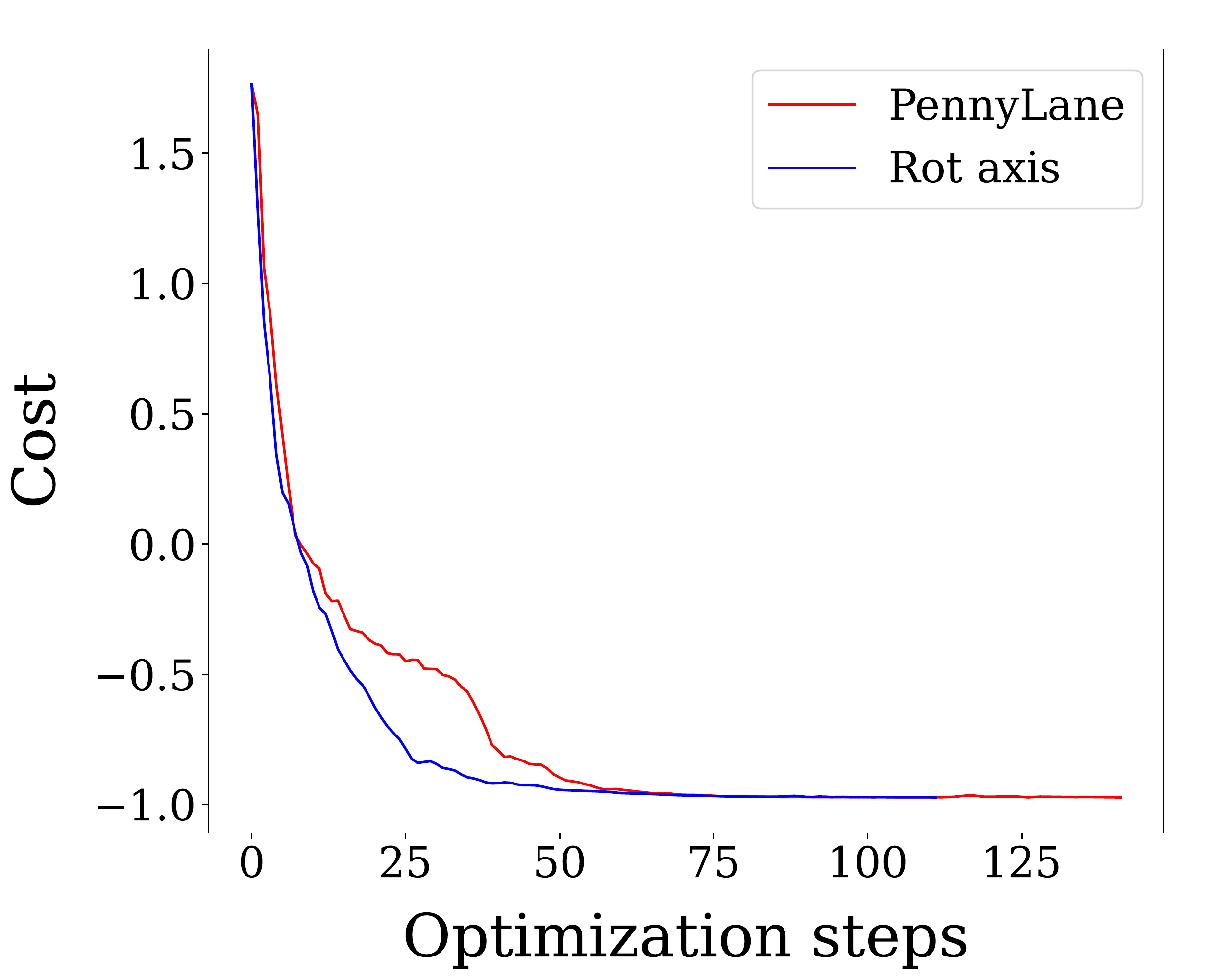}
     \caption {[Color online] Convergence of cost function versus number of optimization steps, for the perturbative gadgets for variational quantum algorithms from Ref. \cite{perturbative_gadget_delay_onset}. Comparison between PennyLane and the rotation implementations.}  
     \label{fig:Pert}
\end{figure}

Additionally, in Ref. \cite{variational_eigenvalue_solver}, the authors use a variational quantum eigensolver in order to find the lowest-energy state of the hydrogen molecule, directly fixing three different spin sectors. In Fig. \ref{fig:Spins} we plot the results for the case of total spin $S=1$, showing a significant advantage of our method both in smoothness and number of iterations to converge to a certain cost value. In order to find the ground state of the molecule our implementation needs 34 iterations, in sharp contrast to the 54 needed by the original method.

Also, in Ref. \cite{perturbative_gadget_delay_onset} the authors proposed to substitute the original Hamiltonian describing our quantum systems/cost function in variational quantum algorithms, by another one called ``gadget" Hamiltonian. The idea is that the gadget Hamiltonian is built from local interactions, so that the landscape of solutions is less likely to present local minima and barren plateaus during the optimization. In this case, our implementation is able to reduce the number of optimization steps from 140 in the original method to 110 by using frame rotations, as shown in Fig. \ref{fig:Pert}.

\begin{figure}
\centering
    \includegraphics[width=\columnwidth]{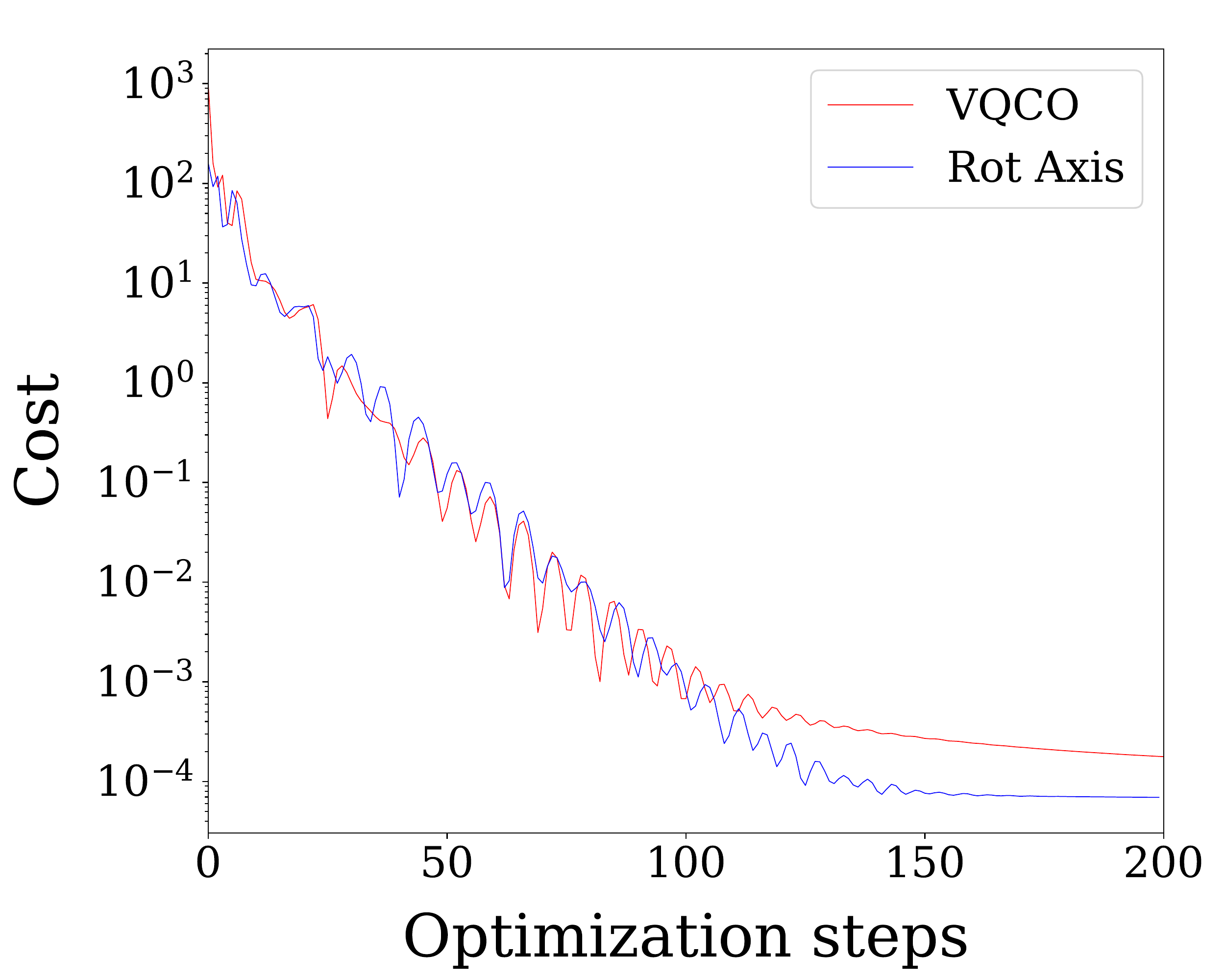}
     \caption {[Color online] Convergence of cost function versus number of optimization steps, for the variational quantum continuous optimizer from Ref. \cite{cv_optimizer}. Comparison between PennyLane and the rotation implementations.} 
     \label{fig:opt8_80}
\end{figure}

Last but not least, in Ref. \cite{cv_optimizer} the authors introduced a variational quantum optimizer to tackle optimization problems with continuous variables, by encoding these variables inside the 3 continuous parameters within each qubit's Bloch sphere. In Fig. \ref{fig:opt8_80} we show how the error in performing the integral of the function $e^{-x^2}$, implemented  by encoding the solution in Fourier modes. We see that the new approach reduces significantly the optimization steps for an arbitrary error bound. 

\bigskip

\section{Conclusions and further work} 
\label{sec4}

In this paper we have proposed a boosting strategy for machine learning methods based on gradient descent, and benchmarked it for different quantum machine learning algorithms. Our idea is based on using rotation matrices involving the parameters of the hyperspace of solutions and the cost and, alternatively, the use of hyperspherical coordinates inside the hyperspace of solutions in which the axis of the cost is also included. The procedure allows to alleviate the effect of many local minima and barren plateaus, thus accelerating the convergence of machine learning methods. 

The procedure discussed in this paper is completely general, and can be applied both to classical and quantum machine learning algorithms. According to our benchmarks, the improvement in convergence, stability, and efficiency can be very significant. As such, we believe that our boosting procedure has a huge potential to improve the performance of AI systems, in turn reducing its computational cost and energy consumption. 

There are different lines of work that can be explored in the future. For instance, it would be interesting to test different changes of coordinates, even dynamical and non-linear ones. Moreover, the performance of the boosting should be further tested in the context of complex deep learning algorithms such as convolutional neural networks, transformers, and generative models, which are computationally quite expensive. In addition, it would also be interesting to explore the performance of this boosting technique in the context of Tensor Network methods \cite{TN}. 

\bigskip 
{\bf Acknowledgements.-} The authors acknowledge DIPC, Ikerbasque, Basque Government and Diputaci\'on de Gipuzkoa for constant support, as well as insightful discussions with the technical teams from Multiverse Computing and DIPC on the algorithms and technical implementations. We also acknowledge Norman Toporcer from Elion for supporting us with several patent applications in connection with parts of the work discussed here.

\bibliography{biblio.bib}

\end{document}